\documentclass[aps,prl,preprint,superscriptaddress,showpacs,showkeys]{revtex4-1}
\usepackage[dvipdf]{graphicx}

\usepackage{color}
\usepackage{bm}
\usepackage{amssymb,amsmath}
\usepackage{color}

\begin{document}

\title{The reversible lithiation of SnO: a three-phase process}

\author{Andreas Pedersen}
\email{andped10@gmail.com}
\author{Petr A. Khomyakov}
\author{Mathieu Luisier}
\affiliation{Integrated Systems Laboratory, Department of Electrical
Engineering and Information Technology, ETH Zurich, Gloriastrasse
35, 8092 Zurich, Switzerland}

\date{\today}

\begin{abstract}


A high reversible capacity is a key feature for any rechargeable battery. In the lithium-ion battery technology, tin-oxide anodes do fulfill this requirement, but a fast loss of capacity hinders 
a
full commercialization. Using first-principles calculations, we propose a microscopic model that sheds light on the reversible lithiation/delithiation of 
SnO and reveals that a sintering of Sn 
causes a strong 
degradation of
SnO-based anodes. When the initial irreversible transformation ends, active anode grains consist of Li-oxide layers separated by Sn bilayers. During the following reversible lithiation, the Li-oxide undergoes two phase transformations that give rise to a Li-enrichment of the oxide and the formation of a layered SnLi composite. We find that the model-predicted anode volume expansion and voltage profile agree well with experiment, and a layered anode grain is highly-conductive and has a theoretical reversible capacity of 4.5 Li atoms per a SnO host unit. The model suggests that the grain structure has to remain layered to sustain its reversible 
capacity and a thin-film design of battery anodes could be a remedy for the capacity loss.

\end{abstract}

\keywords{Tin-oxide, Anode, Reversible, Lithiation, Density Functional Theory}

\maketitle

Tin-based compounds are promising candidates to 
replace graphite as the anode material in lithium-ion batteries
(LIBs)~\cite{Obrovac:2007fw}. Having a maximum capacity of 4.4 Li per host
atom (Li$_{22}$Sn$_5$)~\cite{Boukamp:1981iz}, Sn alloys outperform the
theoretical gravimetric limit of graphite by a factor larger than
two~\cite{Park:2010kj}. 
%
%
The interest for Sn-based material systems was originally sparked by
the pioneering work of Idota {\it et al.}~\cite{Idota:1997co} who
showed that using an amorphous tin composite oxide as the anode
of a LIB cell improves the performance of the battery both in terms of 
capacity and cycleability.

To better understand the behavior of Sn-based materials upon lithiation
and delithiation, Courtney and Dahn~\cite{Courtney:1997wl} conducted
a series of experiments on various Sn-oxides. They found that all
the oxides initially undergo an irreversible Li uptake followed by a regime
where Li insertion and extraction exhibit a reversible behavior. 
Using their experimental findings, Courtney and Dahn developed an empirical
model where it is assumed that an amorphous Li-oxide matrix forms
as the initial Li ions enter the 
pristine
 SnO sample. The O
atoms are captured by Li, which offers stronger bonds as compared to
Sn. At the same time, the Sn atoms
 are subject to a sintering process and form
 clusters embedded
in the emerging Li-oxide matrix. The 
proposed
Sn clustering model is based on
the 
experimental observation of Sn signatures in X-ray
diffraction measurements on the lithiated oxide
materials~\cite{Courtney:1997wl}. Note that the 
aforementioned
structural changes of SnO take place during the initial irreversible
Li uptake. As the lithiation continues, 
the Sn clusters are
assumed to 
behave as a  SnLi$_x$
alloy, where 
$x$ 
is the number of Li atoms per 
Sn, and should provide  
a reversible storage medium for Li.
The model proposed by Courtney and Dahn has served as a
reference for interpreting (de)lithiathion of Li-ion battery
cells with SnO-based anode electrodes. 
We have adapted this model to LiOSn anodes by assuming that the Sn clusters
behave similarly to bulk SnLi$_x$ when exposed to Li.
This will be referred to as the {\it cluster model}.

While the Li-oxide formation and the growth of Sn clusters
 have been confirmed by many
groups~\cite{Chouvin:2000kl,Wang:2000el,Sandu:2004gl,Zhang:2012kq,Jeong:2013eh} 
the actual contribution of these Sn clusters to the reversible lithiation is questionable for several reasons.
Firstly, Li-oxide is an insulator so that it would be difficult
to conceive how electrons can be efficiently 
supplied to the Sn clusters unless
the latter are well interconnected at any stage of lithiation.
%
Secondly, during the initial phase there exist bond types in the LiOSn sample that cannot be explained by the cluster model~\cite{Courtney:1999bt,Wang:2000el,Sandu:2004gl,Sandu:2006ix}.
Thirdly, it is well established that the growth of Sn clusters actually causes a capacity degradation of the anode material~\cite{Courtney:1997br,Behm:2002vk,Jeong:2013eh}.
Fourthly, the initial stress builds up differently in metallic Sn and insulating SnO during the lithiation process~\cite{Tavassol:2014cr} that indicates that the volume expansion profiles are not the same for the two materials.
Finally, a recent experiment by Ebner and co-workers~\cite{Ebner:2013bj} has shown that the
voltage profile and the
volume expansion of SnO oxide during the initial Li insertion/extraction
cycle significantly differ from those of a bulk SnLi alloy.
Fracturing, which was found to occur at 
grain boundaries in SnO, is also difficult to
explain assuming 
a homogenous isotropic amorphous
Li-oxide
 matrix and separated Sn 
 clusters.

\begin{figure*}
\centering
\includegraphics[width=16.cm]{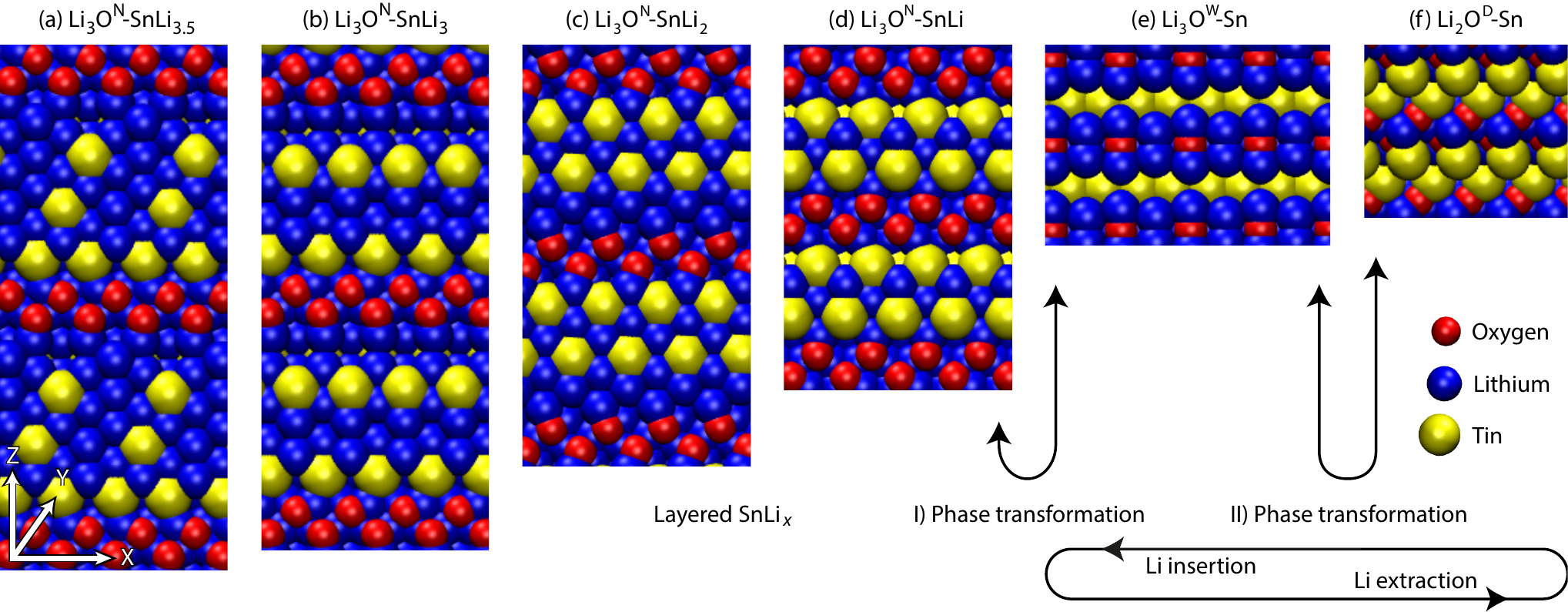}
\caption{Reversible lithiation of SnO. Delithiation is from left depicted by the lowermost arrow. 
Li$_y$O-SnLi$_x$ structural configurations
for $x$ ($y$) decreasing from 3.5 to 0 (from 3 to 2)
and undergoing two 
phase transformations. Blue, red, and yellow
sphere correspond to the Li, O, and Sn atom, respectively.
Initially, 2.5 Li-layers are released,
causing a
Li depletion of
SnLi$_x$ from $x$=3.5 to 1, as shown in
insets~(a)-(d). The remaining Li layer in the SnLi composite
layer is freed as the Li-oxide changes its phase
from Li$_3$O$^\text{N}$ to Li$_3$O$^\text{W}$, inset~(e). Another
phase transformation occurs upon further Li extraction, bringing the Li-rich
oxide to its irreversible Li$_2$O$^\text{D}$ phase,
inset (f).} \label{fig:structures}
\end{figure*}

In this 
paper,
we propose a microscopic
model for reversible (de)lithiation of tin monoxide that resolves 
the shortcomings of the 
cluster
model by Courtney and
Dahn~\cite{Courtney:1997wl} summarized in the previous paragraph. 
Our model is based upon an atomistic study of a large set of structures obtained from first-principles calculations. 
According to our recent {\it ab-initio}
calculations of the irreversible Li-uptake of SnO, a fully Li-depleted anode
consists of layered 
Li-oxide
separated by
Sn bilayers~\cite{Pedersen:2014ko}. Starting from this
configuration, we show here that the Li-oxide undergoes 
two
 phase transformations as the Li concentration increases, giving rise to the formation of 
a Li-rich and layered 
Li-oxide
 (Li$_3$O). Subsequently 
3.5 additional Li atoms per Sn atom are accommodated solely in Sn layers. Thus, the final structure  
contains
6.5 Li per Sn atom in total, but the {\it reversible capacity} of SnO is limited to 
4.5 Li per Sn (3.5) and O (1) host atom whereas the remaining two Li atoms are strongly bound in the Li-oxide.
 By
applying this {\it three-phase model} we are able to reproduce the
experimental data for the volume expansion and the voltage profile measured upon lithiation. We also find that the obtained LiOSn model structures
exhibit 
a good 
in-plane
electron conductance because of their
highly-conductive SnLi$_x$ layers.
The results presented here suggest that the undesired degradation of reversible capacity for Li-ions might be averted by designing an anode that preserves its layered structure during (de)lithiation. This could be achieved by adopting a thin-film structure on a lattice matched substrate to promote the layered structure or by inserting species that retard the ongoing Sn sintering process. These general findings could have a wider applicability and be used to better understand the behavior of other metal oxides~\cite{Zhang:2011hv} or Na-ion batteries~\cite{Kim:2012im}.
%


We now describe in detail the
entire process of the anode (de)lithiation as given by the proposed
three-phase model. The starting configuration corresponds to
the anode grain structure, in which the irreversible uptake of 
Li
 is
fully complete, and the Li-oxide layers are separated by Sn bilayers
as shown in Fig.~\ref{fig:structures}(f).
Layers or half-layers of Li are
inserted into 
the
structure as long as the relative
binding energy for Li remains negative
\begin{align}
\label{eq:energy} E_b &= \frac{E_{L} - \left( E_{L-1}+N_{L} \cdot
E_{\text{Li}} \right)}{N_{L}},
\end{align}
where $E_b$ is the 
average
relative binding energy of a single 
Li atom;
$E_{L}$ is the total energy of the anode structure with $L$ layers of Li; 
$E_{\text{Li}}$ is the cohesive energy of a Li atom in a bulk environment,
and $N_L$ is the number of Li atoms in a single Li layer. Structures
fulfilling the requirement $E_b<0$ in Eq.~(\ref{eq:energy}) have been
determined for Li concentrations up to $L=6.5$.  This Li content
corresponds to a fully lithiated SnO sample. For $L\geq7$ the additional Li atoms are located in the LiOSn structure within an environment similar to
bulk Li and have a comparable binding energy so that $E_b>0$. 
This indicates that eventual overcharging will 
result in the formation of domains with bulk Li, which, in the best
case scenario, just act as passive spectators. 
In
other words, having any extra Li atoms in the fully-loaded anode
structure does not increase the battery capacity since these Li atoms
will not be released by the layered anode 
 upon normal
discharging conditions.   

Figure~\ref{fig:structures} shows all the structural configurations
obtained during the (de)lithiation process described above. 
As the Li load increases, three different phases can be
identified, which are separated by two phase transformations of the Li-oxide.
The first phase, Li$_2$O$^\text{D}$, corresponds to the fully
Li-depleted configuration in Fig.~\ref{fig:structures}(f). The
label $\text{D}$ refers to it being a {\it depleted} structure in which the
metallic Sn bilayers surround the insulating Li-poor oxide layers.
A second phase, Li$_3$O$^\text{W}$, as shown in
Fig.~\ref{fig:structures}(e), appears upon insertion of the first
reversible 
Li layer. 
The XY cross section of this oxide layer expands so
that the label $\text{W}$ stands for its {\it wider} cross section
area. Note that no substantial change of the supercell size occurs in
the Z direction though the Sn bilayer transforms into a monolayer
structure. Four structural configurations of a third Li-oxide phase,
Li$_3$O$^\text{N}$, are shown in Figs.~\ref{fig:structures}(a)-(d).
This 
oxide
results from a
  phase transformation where the number of Li atoms remains unchanged and
 the Li-rich oxide
regains a narrow XY cross section. The label $\text{N}$ refers
to its {\it narrower} cross section area. 
Hereafter, the
inserted Li atoms form a SnLi$_x$ layered composite structure
that is accompanied by a strong expansion of the corresponding
supercell in the Z direction. The layered composite can accommodate
a maximum of 3.5 Li per Sn atom. The fully-loaded structural
configuration is given in Fig.~\ref{fig:structures}(a).
From Fig.~\ref{fig:structures} it is also apparent that the three-phase model offers an explanation for the ``unusual'' bond types
through a high ratio of both 
 Sn-Sn, Sn-Li, Sn-O, and Li-O bonds in the layered anode grains.

\begin{figure}
\centering
\includegraphics[width=8.cm]{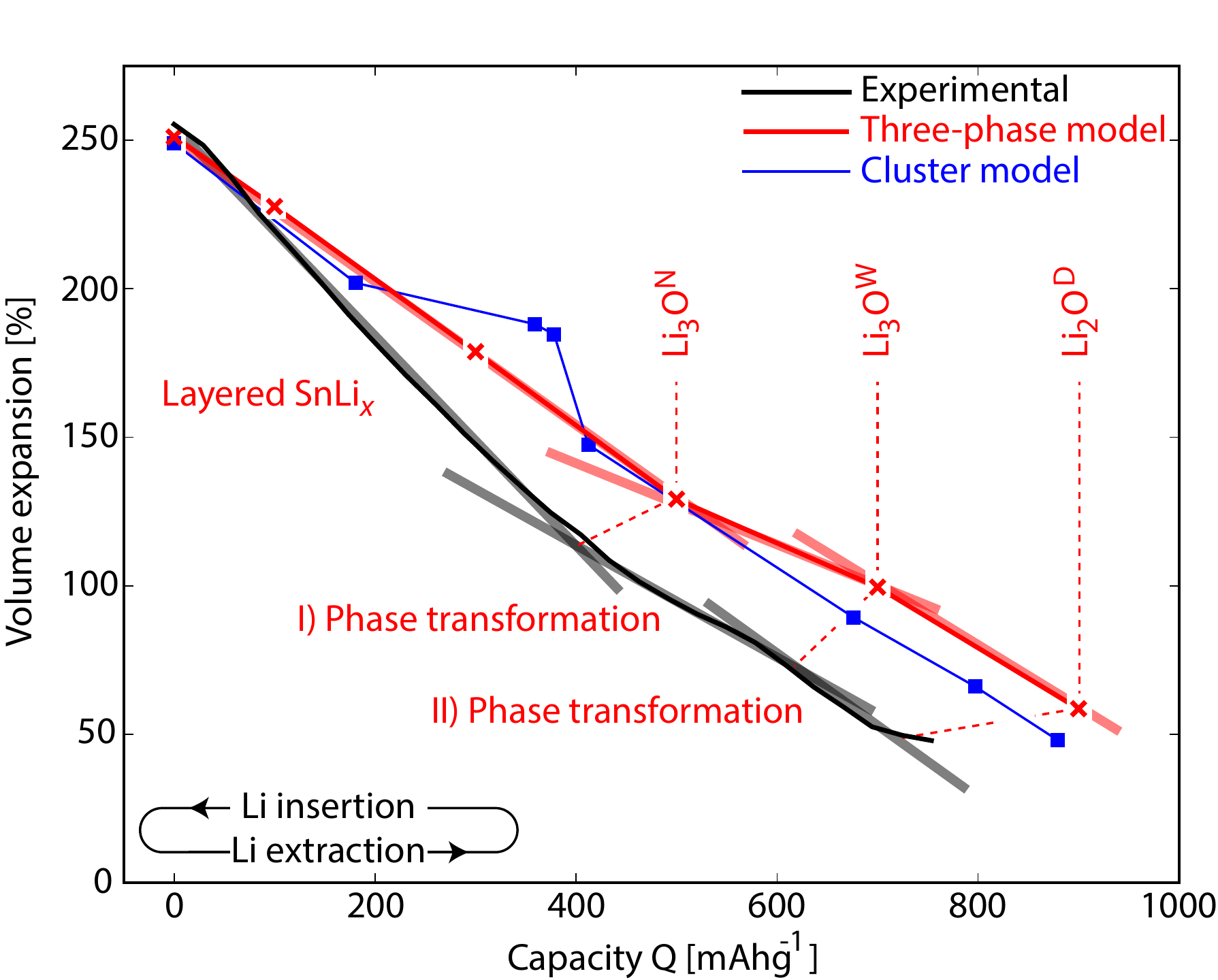}
\caption{Volume expansion as a function of Li content for LiOSn anode. The volume expansion is defined with respect to the
volume of pristine SnO. Red crosses and filled blue squares refer to the
numerical data computed with the three-phase model and the cluster
model, respectively. The black, blue, and red solid lines
correspond to the experimental data~\cite{Ebner:2013bj}, the
cluster, and the proposed three-phase
model, respectively. The solid lines only serve as a guide to the eye.
Three volume expansion regimes can be identified. They are delimited by
the three gray straight lines. The delithiation process
evolves as follows: (i) Li is released from the SnLi$_x$ composite
structure sandwiched between Li$_3$O$^\text{N}$ layers, (ii) the Li-oxide
transforms into Li$_3$O$^\text{W}$ separated by Sn
monolayers, and (iii) a second transformation into the depleted and
layered Li$_2$O$^\text{D}$ oxide with Sn bilayers
in-between occurs.} 
\label{fig:expansion_unscaled}
\end{figure}

In the present study, we only focus on the first reversible
(de)lithiation
 cycle under the assumption that the initial
and irreversible transformation from 
Sn- to Li-oxide
 has already taken
place. 
To validate the three-phase model
against the experimental data in Ref.~\onlinecite{Ebner:2013bj}, we
computed the volume expansion of the Li$_y$O-SnLi$_x$ anode with respect to
the volume of pristine SnO upon Li-extraction, as reported in
Fig.~\ref{fig:expansion_unscaled}. The volume expansion 
is defined as
\begin{align}
\label{equ:volume} \Delta V &=
\frac{V-V_{\text{SnO}}}{V_{\text{SnO}}} \cdot 100\%,
\end{align}
where $V$ is the volume of the Li$_y$O-SnLi$_x$
supercell at a given Li concentration ($x$ and $y$), and $V_{\text{SnO}}$ 
is the volume of a pristine SnO structure with the same number of Sn
atoms as in the corresponding LiOSn supercell.

The three volume expansion regimes, which can be clearly identified in 
Fig.~\ref{fig:expansion_unscaled}, correspond to the three Li-oxide
phases shown in Fig.~\ref{fig:structures}. 
At first, Li atoms are
released from the SnLi$_x$ composite structure, which is
situated in-between Li$_3$O$^\text{N}$ layers.
This causes a 
constant volume reduction imposed by the supercell contraction in the
Z direction, perpendicular to the XY cross section. While
SnLi$_x$ becomes fully Li-depleted, the Li-rich oxide 
undergoes a phase transformation (N$\rightarrow$W) that slows down
the volume reduction. Though the atom rearrangement in the anode grain
structure is rather drastic,
 the corresponding
volume reduction is quite moderate because of the significant XY
cross section expansion, which compensates for the supercell
contraction in the Z direction. Finally, further delithiation
extracts the remaining 
Li atoms
 that are loosely-bound in the oxide,
giving rise to another transformation (W$\rightarrow$L). This brings
back the anode structure into its depleted state Li$_2$O$^\text{D}$ with a narrow
cross section, as depicted in  Fig.~\ref{fig:structures}(f). This
final transformation induces a strong volume reduction.
Figure~\ref{fig:expansion_unscaled} shows that the volume
expansion, as predicted by the three-phase model, is in a good
agreement with the volume change measured in
Ref.~\onlinecite{Ebner:2013bj}. 
We notice that there exists a strong dependence of the anode grain volume on the Li-concentration.
It may give rise to a deterioration of the
layered structure during 
operation if the Li concentration is
inhomogeneous over the anode volume. This could explain
why Li-oxide has been classified as amorphous in LiOSn anodes~\cite{Courtney:1997wl}.
It also explains why fracturing tends to occur at grain boundaries. 
These fractures take place to relieve accumulated strain in the regions where crystal grains of different orientation and shape intersect during volume expansion and contraction.

\begin{figure}
\centering
\includegraphics[width=8.cm]{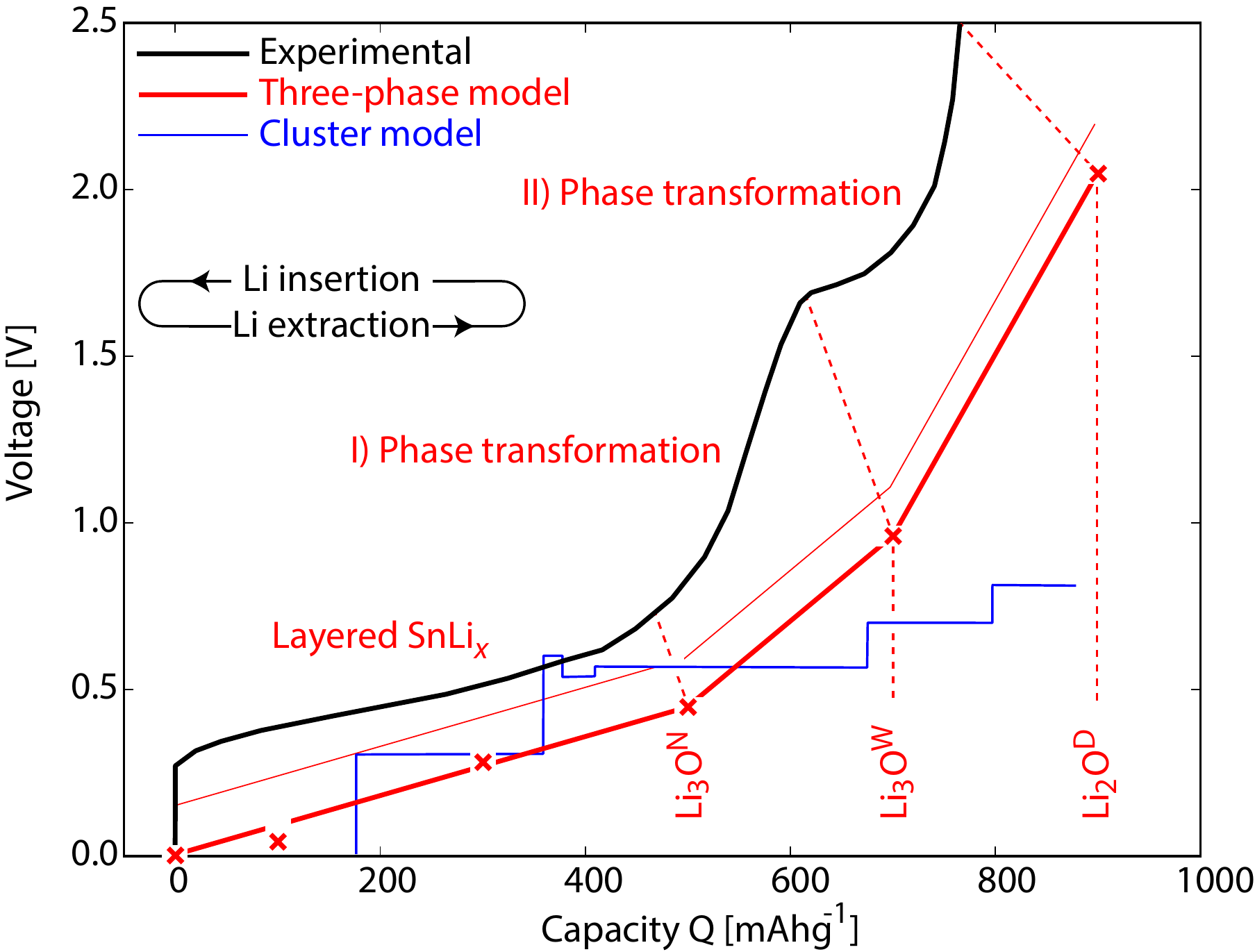}
\caption{ Voltage profile as a function of Li content for LiOSn anode. The voltage is defined with respect
  to a Li counter electrode (zero-voltage reference). The same
  plotting conventions as in Fig.~\ref{fig:expansion_unscaled} are
  used, thin red line is shifted by $V_\text{o}$ (0.16~V). The data points for a bulk SnLi$_x$ alloy are shifted along the X-axis
  to coincide with the corresponding points in
  Fig.~\ref{fig:expansion_unscaled}. The first step-like segment of
  the voltage profile refers to the SnLi$_x$ delithiation. The
  two segments with a steeper slope correlate with a phase
  transformation and 
  a
  Li-depletion of the Li-oxide, first from
  Li$_3$O$^\text{N}$ to Li$_3$O$^\text{W}$, followed by the phase transformation
  to Li$_2$O$^\text{D}$.}
   \label{fig:voltage_both}
\end{figure}

 To  
further
 validate the
three-phase model, we calculated the voltage profile of the
LiOSn anode during (de)lithiation, as
shown in Fig.~\ref{fig:voltage_both}. 
A
previous work on bulk SnLi alloys~\cite{Courtney:1998bw} demonstrated
that the voltage profile, which is determined by the change of the
Gibbs free energy, can be obtained from the internal
energy change only, while the entropy and volume terms can be
safely neglected. We adopted a similar approach to compute the voltage
profile during delithiation
\begin{align}
\label{equ:voltage} \Delta U &= \frac{\left( E_{6.5-\Gamma} + \Gamma
\cdot N_{L} \cdot E_{\text{Li}} \right) - E_{6.5}}{\Gamma \cdot
N_{L}\cdot |e|},
\end{align}
where $\Delta U$ is the voltage change relative to the
fully-lithiated anode structure with the total energy $E_{6.5}$;
$E_{6.5-\Gamma}$ is the total energy of an anode structure with
$\Gamma$ Li layers removed; $N_L$ is the number of Li atoms in a
single Li layer, and $e$ is the elementary charge of an electron.

Figure~\ref{fig:voltage_both} shows that a rather moderate
increase of the voltage takes place as the Li atoms are released
from the SnLi$_x$ composite structure during the first stage of
delithiation ($0<Q<500$~mAhg$^{-1}$).
A steep voltage increase then occurs as the SnLi composite is depleted of Li atoms and the Li-rich oxide undergoes a phase transformation 
($500<Q<700$~mAhg$^{-1}$).
Finally, the further extraction of Li atoms from the Li-rich oxide ($700<Q<900$~mAhg$^{-1}$) increases the voltage up to its maximum value of 2.05~eV for the fully
Li-depleted anode structure (Fig.~\ref{fig:structures}(f)). 
One can
see in Fig.~\ref{fig:voltage_both} that the three-phase model
reproduces the experimentally observed voltage profile
in a semiquantitative manner~\cite{Ebner:2013bj,Courtney:1997wl, Chouvin:2000kl, Wang:2000el}. 
Figure~\ref{fig:voltage_both} also show that the behavior of a bulk SnLi alloy and lithiated SnO oxide is qualitatively different, which  demonstrates that the Sn cluster model is not applicable for the LiOSn anode grain.

We notice that the voltage profile given by the three-phase model
misses the 
onset of~0.3~V that is present in the
experimental data at the capacity $Q=0$. This can be explained by
the fact that the computed voltage is determined from an average
change in the total energy of the anode, as given in
Eq.~(\ref{equ:voltage}), whereas eventual activation energy 
barriers of the Li atom diffusion process upon delithiation are not accounted for. 
As 
a first approximation,
the onset voltage can be estimated in the following way.
When removing a single Li atom from the anode structure, it
appears that the Li atoms residing at the interface between the
Li-oxide
 and 
 the composite
 layers have the weakest bonds. This holds true
for all the structures, except the one with a fully Li-loaded anode
(Fig.~\ref{fig:structures}(f)) in which the least bound Li is in the
SnLi composite layer. The energy required to extract this Li atom
from the composite layered structure is 0.16~eV, i.e., 0.16~V can be  
considered as the lower bound for the onset voltage ($V_\text{o}$) at the capacity
$Q=0$.

The anode grain structures, which are given by the
three-phase model, have been assumed to be good electrical conductors
during the entire (de)lithiation cycle. This means that all
the layered structures in Figs.~\ref{fig:structures}(a-f)
should have a high electron conductance in the XY-plane being 
comparable to
that of bulk $\beta$-Sn and Li. To confirm this hypothesis, we have
calculated the band structure of the Li-oxide and SnLi composite
layers, which reveals that the oxide layer is an insulator with a
sizable band gap that hinders electron transport in the Z-direction, whereas the SnLi composite layer is metallic and
provides conducting channels for in-plane electron transport. First-principles
calculations of the ballistic electron transport through
the SnLi composite layers in Fig.~\ref{fig:structures} explicitly show 
that the in-plane electron conductance is comparable to that of bulk
$\beta$-Sn and Li calculated using the same approach. Hence, the
proposed anode structure is a good conductor, which is
consistent with experimental observations. Details about the 
conductance calculations will be published elsewhere
\cite{Khomyakov:unpublished}.

We have done first-principles,
density-functional theory calculations to understand the reversible 
lithiation/delithiation of SnO anodes in Li-ion batteries at the
microscopic level. Based on these atomistic simulations, we developed
a three-phase model that consistently explains how the 
 structural and
electronic properties of a SnO-based anode (phase transformation, volume expansion, voltage
profile, and electron conductance) evolve during the initial
(de)lithiation cycle. 
This model predicts
that the reversible capacity of the SnO-based Li-ion battery can reach
up to 4.5 Li atoms per SnO host unit, which is slightly higher than the
theoretical capacity for bulk Sn~\cite{Boukamp:1981iz}.
It also sheds light onto the 
surprising
experimental observation of ``unusual'' bonds that have been explained by a high ratio of Sn-Sn, Sn-Li, Sn-O, and Li-O bonds in the layered LiOSn structure.
Using the obtained results, we proposed a plausible 
explanation for the deterioration and amorphization of the LiOSn structure, which 
are attributed to spacial
inhomogeneities of the Li concentration in real samples.
Finally, the observed capacity degradation of SnO anodes can be understood by applying
the cluster model by Courtney and Dahn~\cite{Courtney:1997wl} to describe
the slower but irreversible structural changes.
A loss in Li capacity results from the system transformation 
towards its thermodynamically more
 stable configuration with Sn clusters embedded in a lithium oxide rather than remaining layered as required by the reversible three-phase model. 

Our findings suggest that improved SnO-based anodes could be achieved 
by applying a thin-film design or use an additive to retard the agglomeration of Sn. 
Furthermore, the layered character of the three-phase model structures appears to be  general and might be used to understand the lithiation process of other metal oxides. This is supported by a similarity of voltage profiles measured in Refs.~\onlinecite{Poizot:2000du,Poizot:2002fl} for the lithiation of transition metal oxides. The model might also shed light onto the charging dynamics of sodium-ion batteries~\cite{Su:2013ev, Han:2014ju}, where Na-ions replace Li-ions as the charge carriers.

\section{Methods}
Our first-principles calculations rely on a density-functional plane-wave pseudo-potential method within
the framework of the generalized gradient approximation
(GGA-PBE)~\cite{Perdew:1996ug} and the projector augmented wave
(PAW) formalism~\cite{Blochl:1994uk}, as implemented in the VASP
code~\cite{Kresse:1996kl,Kresse:1996vk}. The equilibrium
LiOSn structures consist of a 8x8x1 SnO supercell (64 Sn and 64 O atoms) with  
a Li content ranging from 128 to 448 atoms.
The supercell Brillouin zone is sampled with a
2x2x2 k-point grid. The plane-wave kinetic energy cut-off is set at
500~eV. The total energy and forces are converged to at least
$10^{-8}$~eV and $10^{-3}$~eV/\AA, respectively. Further computational details can be found elsewhere~\cite{Pedersen:2014ko}.

\section{Acknowledgments}
This research is funded by the EU Commission (the ERC starting
grant: E-MOBILE). The computer simulations are done at the Swiss
National Supercomputer Center (project: s579). The authors thank
Martin Ebner for helpful discussions.

\bibliographystyle{apsrev4-1}
\bibliography{Li2OSn_lithiate}

\end{document}